\documentclass[preprint,showpacs,showkeys,amsmath,amssymb]{revtex4}
\usepackage{graphicx}
\usepackage{dcolumn}
\usepackage{bm}

\newcommand\onevec{\ensuremath{\mathbf{1}}}
\newcommand\onevect{\ensuremath{\mathbf{1}^T}}
\newcommand\I{\ensuremath{\mathbf{I}}}
\newcommand\G{\ensuremath{\mathbf{G}}}
\newcommand\Lap{\ensuremath{\mathbf{L}}}
\newcommand\dvec{\ensuremath{\mathbf{d}}}
\newcommand\dvect{\ensuremath{\mathbf{d}^T}}
\newcommand\D{\ensuremath{\mathbf{D}}}
\newcommand\Linv{\ensuremath{\mathbf{L}^{+}}}
\newcommand\uvec{\ensuremath{\mathbf{u}}}

\newcommand\mt{\ensuremath{\tilde{\mu}}}
\newcommand\dbar{\ensuremath{\overline{d}}}

\newcommand\uvecz[1]{\ensuremath{\mathbf{u}^{(0,#1)}}}
\newcommand\lamz[1]{\ensuremath{\lambda_0^{(#1)}}}
\newcommand\be{\begin{equation}}
\newcommand\te{\end{equation}}
\newcommand\bea{\begin{align}}
\newcommand\eea{\end{align}}
\newcommand\pvec{\ensuremath{\mathbf{p}}}
\newcommand\qvec{\ensuremath{\mathbf{q}}}
\newcommand\tti{\ensuremath{\tilde{t}}}

\begin{document}


\title{Molecular Clock on a Neutral Network}

\author{Alpan Raval}

 \email{araval@kgi.edu}
\affiliation{%
Keck Graduate Institute of Applied Life Sciences, 535 Watson Drive, Claremont, California 91711, USA\\
School of Mathematical Sciences, Claremont Graduate University, 711 N. College Avenue, Claremont, California 91711, USA
}%

\date{\today}

\begin{abstract}
The number of fixed mutations accumulated in an evolving population 
often displays a variance that is significantly larger than the 
mean (the overdispersed molecular clock). 
By examining a generic 
evolutionary process on a neutral network of high-fitness genotypes, 
we establish a formalism 
for computing all cumulants of the full probability 
distribution of accumulated mutations in terms of graph
properties of the neutral network, and use the formalism to prove 
overdispersion of the molecular clock. We further 
show that significant overdispersion arises naturally in evolution
when the neutral network is highly sparse, exhibits 
large global fluctuations in neutrality, and small 
local fluctuations in neutrality. The results are also relevant for
 elucidating 
the topological structure of a neutral network from 
empirical measurements of the substitution process.
\end{abstract}

\maketitle

{\it Introduction}. -- The neutral theory of molecular evolution \cite{kimura83} posits that most sequence substitutions at the nucleic acid or 
protein level are selectively neutral and do not 
appreciably alter the activity of the molecule in which they 
occur or the fitness of the host organism. It predicts that 
the number of substitutions accumulated in an
evolving population of sequences in time $t$ follows a Poisson distribution
with mean $\mu \nu t$, where $\mu$ is the mutation rate per sequence per generation 
and $\nu$ is the average fraction of neutral mutations (also called the
neutrality). This prediction gives a simple explanation to the 
``molecular clock'' \cite{zucker} -- the idea that
the number of accumulated fixed mutations in a population is proportional to
the time elapsed -- and implies that the variance in
this number must equal its mean, leading to an
index of dispersion (defined as the variance divided by the mean) of $1$.

However, experimental 
studies often find that the index of dispersion
is significantly larger than $1$
(the overdispersed molecular clock) \cite{ohta71,langley74,gillespie84}
This finding can be reconciled with the neutral theory
by assuming that the space of neutral sequences has fluctuating
neutrality \cite{takahata87}, causing the substitution 
process to be 
non-Poissonian, as verified 
by computer simulations
 \cite{bastolla99,bastolla2002,wilke2004} 
that show significant overdispersion when the product 
$N\mu$ of the 
population size $N$ and the mutation rate $\mu$ is much smaller than $1$. 

There is limited theoretical understanding of the nature
of the molecular clock. Cutler \cite{cutler2000a} formally 
calculated the index of dispersion in terms of 
statistics of the mutation and fixation processes
and argued that slow
fluctuations in evolutionary parameters could lead to significant 
overdispersion in simple evolutionary models. Recent analytical results 
include a derivation of the
index of dispersion for neutrally evolving protein populations constrained
by a stability requirement \cite{bloom2007}. These results do not 
conclusively prove overdispersion of the molecular clock 
in a sufficiently general 
scenario, nor
do they give an explicit characterization of the non-Poissonian nature of
the full probability distribution of accumulated mutations.

A natural stage for fluctuating neutrality is presented by a neutral network \cite{smith70,huynen96,govindarajan97,bornberg99,nimwegen99,tiana2000,bastolla2002}
of high- and equal-fitness genotypes in which two genotypes are linked by an
edge if they differ by a single point mutation. 
The aim of this Letter is to theoretically 
clarify the non-Poissonian nature of the distribution of 
accumulated fixed mutations, to relate all 
cumulants of this distribution to graph invariants of the 
neutral network, to prove overdispersion of 
the molecular clock, and to identify features of the neutral network that 
could lead to significant overdispersion. 
We assume $N\mu \ll 1$, as is relevant for the majority of organisms in the plant and animal kingdom \cite{wilke2004}. 
For this limit to be valid, it is also necessary that $\mu \ll 1$, 
as we assume below. We further assume
that the neutral network is a connected graph; if it is not connected, the 
results below apply separately to populations evolving on each connected 
component of the neutral network graph.

{\it Substitution process when $N\mu \ll 1$}. -- Consider a population of 
$N$ individuals evolving on a neutral network, represented by a graph 
$\mathfrak{G}$ with $n$ nodes, $E$ edges, and adjacency matrix $\G$. 
The nodes of 
$\mathfrak{G}$ represent high-fitness genotypes 
characterized by sequences of length $L$ over an alphabet of size $A$. Two
nodes are connected by an edge if the corresponding genotypes differ by a
 single 
point mutation.  The neutrality of a node $r$ in $\mathfrak{G}$ 
is $d_r/(L(A-1))$, where $d_r$ is the degree of $r$ in $\mathfrak{G}$, 
and represents the fraction of point mutations of the genotype $r$ that 
are neutral. Following \cite{nimwegen99}, we consider a discrete 
mutation-selection
dynamics in which at each generation an individual suffers a point mutation with fixed 
probability $\mu$ that moves it to a neighboring genotype (which may or may not be of high
fitness). $N$ individuals are then selected with replacement 
from the mutated 
population with 
probability proportional to their fitness, and the process is repeated. 
For $N\mu \ll 1$, 
the population at any point in 
time is converged on a single node of the neutral network \cite{nimwegen99}. 
At each generation it either stays at its current node or moves effectively 
as a single entity 
to a neighboring node. The probability
$p_t(r)$ that the population is on node $r$ at time $t$ is governed by the equation \cite{bloom2007,nimwegen99}
\be
\label{vn2}
\pvec_t = \left(\I-\mt \D + \mt \G \right)\pvec_{t-1} = \left(\I-\mt \Lap \right)\pvec_{t-1},
\te
where $p_t(r)$ is the $r$th element of $\pvec_t$, $\I$ is the $n \times n$ identity matrix, $\mt \equiv \mu/(L(A-1))$ is the reduced mutation rate, $\D$ is 
a diagonal matrix with node degrees on the main diagonal, and 
$\Lap \equiv \D-\G$ is the graph Laplacian of $\mathfrak{G}$. The term $\I -\mt \D$ represents the probability that the population stays at its current
node (either due to no mutation or a deleterious mutation 
that is culled by selection), and the term $\mt \G$ represents the 
probability that the population moves to a neighboring node. 
 
%
$\Lap$ is a symmetric, 
positive semi-definite matrix and, 
if $\mathfrak{G}$ is connected,
has exactly one zero eigenvalue and all other eigenvalues positive \cite{biggs76,chung94}. 
We denote eigenvalues 
of $\Lap$ by $\lambda_0 < \lambda_1 \leq \lambda_2 \leq \ldots \leq \lambda_{n-1}$, with
$\lambda_0 = 0$. Further, because $\sum_j L_{ij}=\sum_i L_{ij} = 0$, the eigenvector of $\Lap$ 
corresponding to $\lambda_0$ is proportional to \onevec , the column vector with all entries equal to $1$. 
The properly normalized limiting distribution over $\mathfrak{G}$ is $\lim\limits_{t \rightarrow \infty}\pvec_t = n^{-1}\onevec$,
i.e., all nodes are occupied with equal probability \cite{nimwegen99}. 

Consider the joint distribution
$p_t(r,m)$, representing the probability that the population is on node $r$ at time $t$ and $m$ 
neutral substitutions have accumulated since time $0$ (see \cite{bloom2007} for a similar representation). 
The dynamics of the joint process is 
\be
\label{fundeq}
\pvec_t(m)=\left(\I-\mt \D\right)\pvec_{t-1}(m)+\mt \G\pvec_{t-1}(m-1),
\te
where the $r$th element of $\pvec_t(m)$ is $p_t(r,m)$. Assuming an 
equilibrated population at $t=0$, the initial condition for
Eq. (\ref{fundeq}) is 
$\pvec_0(m)=\delta_{m,0}n^{-1}\onevec$. 

To solve (\ref{fundeq}), it is convenient to define
the vector moment generating function (mgf) $\qvec_t(\theta)=\sum\limits_{m=0}^{\infty}e^{m\theta}\pvec_t(m)$. Noting that $\qvec_0(\theta)=n^{-1}\onevec$, 
multiplying both sides of Eq. (\ref{fundeq}) by $e^{m\theta}$, summing over all $m$, and finally solving the resulting equation yields
\be
\label{vmgfsoln}
\qvec_t(\theta)=n^{-1}\left(\I-\mt \Lap +\mt (e^{\theta} -1)\G\right)^t \onevec.
\te
The mgf $q_t(\theta)$ for the distribution of accumulated mutations is found
by marginalizing over the vector mgf: $q_t(\theta)=\sum_r q_t(r,\theta) = \onevect \qvec_t(\theta)$, where the superscript $T$ 
denotes the transpose operation. This yields 
\be
\label{exactmgf}
q_t(\theta)=n^{-1}\onevect \left(\I-\mt \Lap +\mt (e^{\theta} -1)\G\right)^t \onevec.
\te
The probability $p_t(m)$ that $m$ mutations have accumulated in time $t$ may be
recovered as the coefficient of $e^{m\theta}$ in the above mgf. 
%
Moments of $p_t(m)$ are 
obtained in the usual manner by taking multiple derivatives of 
 Eq. (\ref{exactmgf})
with respect to $\theta$. This procedure, however, becomes increasingly 
cumbersome for the calculation of higher moments, primarily because 
$\Lap$ and $\G$ do not, in general, commute. We therefore directly consider the late time and small
$\mt$ limit of the mgf $q_t(\theta)$ below.

{\it Late time behavior and cumulants}. -- 
Consider a time scale long 
enough so that a sufficiently large number of mutations have accumulated in 
the population, i.e., $t \gg \mt^{-1}$. It is then convenient to 
measure time in units of $\mt^{-1}$, define a rescaled time variable 
$\tti = \mt t$, and examine Eq. (\ref{exactmgf}) in the limit $\tti \gg 1$ 
and $\mt \ll 1$. Equation (\ref{exactmgf}) may be rewritten as
\bea
q_t(\theta) &= n^{-1}\onevect \left(\I-\mt \Lap +\mt (e^{\theta} -1)\G\right)^{\tti/\mt} \onevec,\\
&\simeq n^{-1}\onevect \exp\left[\tti \left((e^{\theta} -1)\G-\Lap \right)\right] \onevec, \label{approxmgf}
\end{align}
where we have used $\mt \ll 1$ in making the approximation above. We now 
introduce 
the spectral expansion
\be
\label{spec}
(e^{\theta}-1)\G -\Lap = \sum_{i=0}^{n-1}\lambda_i(\theta)\uvec^{(i)}(\theta){\uvec^{(i)}}^T(\theta),
\te
where $\{\uvec^{(i)}(\theta)\}$ is an orthonormal basis of eigenvectors of $(e^{\theta}-1)\G-\Lap$ with eigenvalues $\lambda_i(\theta)$ ordered in decreasing
order. Note that $\lambda_i(0)=-\lambda_i$ (eigenvalues of $-\Lap$), and in particular, $\lambda_0(0)=0$ and $\uvec^{(0)}(0)=n^{-1/2}\onevec$. For large $\tti$, 
Eq. (\ref{approxmgf}) is dominated by the leading term in Eq. (\ref{spec}), 
corresponding to the largest eigenvalue $\lambda_0(\theta)$. Thus, in the late time limit, the
cumulant generating function $\ln q_t(\theta)$ and associated cumulants 
$\{k_{(j)}\}$ are given by
\be
\ln q_t(\theta) \simeq \tti \lambda_0(\theta),~~~k_{(j)}\simeq \tti \frac{d^j}{d\theta^j}\lambda_0(\theta)\vert_{\theta=0}.
\te
%
Since $\lambda_0(\theta)$ only depends on the topology of the neutral network
graph, it follows that the ratio of any 2 cumulants depends only on the 
topology of the neutral network graph, and not on $\mt$ and $t$, at late times.

To obtain explicit formulae for the cumulants, we need to 
find $\lambda_0(\theta)$ to any desired order in powers of $\theta$. This 
is carried out in a recursive manner: expand $\lambda_0(\theta)$ and 
$\uvec^{(0)}(\theta)$ in  power series in $\theta$,
\be
\label{powser}
\lambda_0(\theta)=\sum_{j=0}^{\infty}\lambda_0^{(j)}\theta^j,~~~\uvec^{(0)}(\theta)=\sum_{j=0}^{\infty}\uvecz{j}\theta^j,
\te
substitute these expansions in the eigenvalue equation
\be
\label{eigeneq}
\left[(e^{\theta}-1)\G - \Lap\right]\uvec^{(0)}(\theta)=\lambda_0(\theta)\uvec^{(0)}(\theta),
\te
and compare the coefficients of equal powers of $\theta$ on both sides of the 
above equation. Noting that $\lambda_0^{(0)}=0$, comparison of coefficients of 
$\theta^0$ on both sides of Eq. (\ref{eigeneq}) yields $\uvecz{0}=n^{-1/2}\onevec$,
and for $j>0$,
\be
\label{fundrec}
\Lap \uvecz{j} = \frac{1}{\sqrt{n}}\left[\frac{\dvec}{j!} -\lamz{j}\onevec\right]+\sum_{l=1}^{j-1}\left[\frac{\G}{(j-l)!}-\lamz{j-l}\I\right]\uvecz{l},
\te
where $\dvec$ is a column vector containing node degrees (the main diagonal of
 $\D$), and it is understood that the sum on the right hand side vanishes
for $j=1$. Multiplying both sides of Eq. (\ref{fundrec}) by $\onevect$ and
using $\onevect \Lap = 0$, one obtains, for $j>0$,
\be
\label{lamrec}
\lamz{j}=\frac{\dbar}{j!}+\frac{1}{\sqrt{n}}\sum_{l=1}^{j-1}\left[\frac{1}{(j-l)!}\dvect - \lamz{j-l}\onevect \right]\uvecz{l},
\te
where $\dbar=n^{-1}\sum_r d_r=2n^{-1}E$ is the average degree of $\mathfrak{G}$. Equation (\ref{lamrec}) recursively expresses $\lamz{j}$ in terms of $\lamz{k}$ and $\uvecz{k}$ 
for $k<j$. To find $\uvecz{k}$, one may consider inverting $\Lap$
in Eq. (\ref{fundrec}). However $\Lap$, since it has a zero eigenvalue, has 
no inverse. We therefore introduce a pseudo-inverse of $\Lap $, denoted 
$\Linv$, and defined by the spectral expansion
\be
\label{pseudo}
\Linv = \sum_{i=1}^{n-1}\lambda_i^{-1}\uvec^{(i)}(0){\uvec^{(i)}}^T(0).
\te
Note that we have omitted the zero eigenvalue in carrying out the 
inversion. $\Linv$ is a positive semi-definite symmetric matrix with $\onevect \Linv = \Linv \onevec = 0$. Equation (\ref{fundrec}) may now be solved by 
writing $\uvecz{j}$ as $\Linv$ multiplying the right hand side plus an 
arbitrary vector in the null space of $\Lap$. However, since $\Lap$ has 
only a single zero eigenvalue, this null space is 1-dimensional. Further,
$\onevec$ lies in this null space; the null space is therefore spanned by 
$\onevec$, and the solution to Eq. (\ref{fundrec}) is
\bea
\label{urec}
\uvecz{j}=&\frac{\Linv \dvec}{j!\sqrt{n}} + \sum_{l=1}^{j-1}\Linv \left(\frac{\G}{(r-j)!} - \lamz{j-l} \I \right)\uvecz{l}\nonumber \\
&-\frac{1}{2\sqrt{n}}\left(\sum_{l=1}^{j-1}{\uvecz{j-l}}^T \uvecz{l}\right)\onevec.
\end{align}
where the coefficient multiplying $\onevec$ above is found, after 
some algebra, by expanding the 
normalization condition ${\uvec^{(0)}(\theta)}^T \uvec^{(0)}(\theta)=1$ 
in powers of $\theta$.

Equations (\ref{lamrec}) and (\ref{urec}), together with the starting 
conditions $\lamz{0}=0$ and $\uvecz{0}=n^{-1/2}\onevec$, are coupled 
nonlinear equations that allow one to recursively find $\lamz{j}$ (and therefore $k_{(j)}$) for all $j$. For example, using these equations and 
noting that $k_{(j)}=\tti j! \lamz{j}$, the first three cumulants at late
times are obtained as
\bea
\label{mean}
k_{(1)}=& \tti \dbar, \\
\label{var}
k_{(2)}=& \tti \dbar + \frac{2\tti}{n}\dvect \Linv \dvec, \\
k_{(3)}=& \tti \dbar + \frac{6\tti}{n}\left[\dvect \Linv \dvec - 2\dbar\, \dvect {\Linv}^2 \dvec+\dvect \Linv \G \Linv \dvec\right] \nonumber.
\end{align}
Since a Poisson distribution with the same mean has all cumulants equal to $\tti \dbar$ (obtained from the first term in Eq. (\ref{lamrec})), Eq. (\ref{lamrec}) shows systematic departures from Poissonian behavior at all cumulant 
orders in a manner that depends purely on the topology of the neutral 
network graph. Further, since for large $\tti$, the Poisson distribution 
may be well approximated by a Normal distribution, the cumulants may be 
used to develop 
an Edgeworth expansion of $p_t(m)$ around
a Normal distribution to any desired accuracy. Fitting this distribution 
to an empirically obtained $p_t(m)$ distribution should then yield finer aspects of
the topology of the neutral network than is accessible from mutational 
robustness studies alone \cite{nimwegen99,bloom2008}.   

{\it Overdispersion of the molecular clock}.-- Since the 
first cumulant is the mean and the second cumulant the variance, the index of 
dispersion $R$ may be found as the ratio of $k_{(2)}$ and $k_{(1)}$ from 
Eqs. (\ref{mean}) and (\ref{var}):
\be
\label{iod}
R = 1+\frac{2}{n\dbar}\dvect \Linv \dvec.
\te
Because $\dvect \Linv \dvec$ is a quadratic form associated with a positive 
semi-definite matrix \cite{elecres}, this shows that 
the molecular clock is generically 
overdispersed ($R\geq 1$). $R=1$ only if the neutral network 
graph is regular because
for a regular graph (and only a regular graph), 
$\dvec \propto \onevec$ lies in the null space of $\Lap $ and $\Linv $. 
Using Eq. (\ref{approxmgf}), it is in fact trivial to 
show that the substitution process is strictly Poissonian for regular 
neutral network graphs, since $\G$ and $\Lap$ commute for regular graphs. 
For all other graphs, $\dvec $ will have a 
component orthogonal to the 
null space of $\Lap $ and thus result in overdispersion. 
This is consistent with having ``fluctuating neutrality'', i.e., unequal neutrality 
across the network, for overdispersion. To examine how the extent of 
overdispersion depends on neutrality fluctuations and other graph parameters,
we now determine bounds on $R$.

Using the spectral expansion (\ref{pseudo}), we obtain
\bea
\label{specquad}
\dvect \Linv \dvec &= \sum_{i=1}^{n-1} \lambda_i^{-1} \left(\dvect \uvec^{(i)}(0)\right)^2, \\
& \leq \lambda_1^{-1}\sum_{i=1}^{n-1} \left(\dvect \uvec^{(i)}(0)\right)^2 = n\lambda_1^{-1} \text{Var}(d),
\end{align}
where we have used the fact that $\lambda_1$ is the second-smallest eigenvalue of $\Lap$ and that $\{\uvec^{(i)}(0)\}$ is an orthonormal basis of eigenvectors. $\text{Var}(d)$ denotes the variance of the degree distribution of the graph. Noting that $2/(n\dbar)=1/E$, this results in an upper bound on the extent of overdispersion:
\be
\label{rupper}
R -1\leq  \left(\frac{n}{E}\right)\lambda_1^{-1} \text{Var}(d).
\te
Thus the index of dispersion is bounded from above by an 
interesting combination of graph parameters: the sparseness (as measured by 
the ratio $E/n$), the fluctuations in neutrality (as measured by $\text{Var}(d)$), and $\lambda_1^{-1}$, which has a number of interpretations. $\lambda_1$ is
the algebraic connectivity of the graph \cite{fiedler} and measures its 
overall compactness and connectivity. Also, $\lambda_1^{-1}$ is the time
scale (as measured in units of $1/\mt$) of relaxation of the distribution 
$\pvec_t$ (Eq. (\ref{vn2})) to its equilibrium value $n^{-1}\onevec$. 
Therefore, for a 
fixed amount of neutrality fluctuation, $R$ can be large if the neutral
network is sparse (high $n/E$) and less well connected, or equivalently, 
if the network is sparse and the relaxation time scale is large. 
Since both 
of these conditions are expected to hold quite generally 
for large and sparse neutral 
networks, Eq. (\ref{rupper}) is a weak upper bound. It is more 
interesting to examine the following lower bound on $R$. Returning to Eq. (\ref{specquad}), and using the fact that $f(\lambda_i) \equiv \lambda_i^{-1}$ is
a convex function of $\lambda_i$ for positive $\lambda_i$, we may apply Jensen's inequality for convex functions:
\be
\frac{\sum_i a_i f(\lambda_i)}{\sum_i a_i} \geq f\left(\frac{\sum_i a_i \lambda_i}{\sum_i a_i}\right)
\te
to Eq. (\ref{specquad}) with the choice $a_i = \left(\dvect \uvec^{(i)}(0)\right)^2$. This yields a lower bound on $R$, namely
\bea
R-1 &\geq \frac{n^2}{E}\frac{\text{Var}(d)^2}{\sum_{i=1}^{n-1} \lambda_i \left(\dvect \uvec^{(i)}(0)\right)^2}\nonumber \\
&=2\left(\frac{n}{E}\right)^2 \frac{\text{Var}(d)^2}{\sum_{i,j}G_{ij}\left(d_i-d_j\right)^2},
\end{align}
where we have used $\sum_{i=1}^{n-1} \lambda_i \left(\dvect \uvec^{(i)}(0)\right)^2=\dvect \Lap \dvec=(1/2)\sum_{i,j}G_{ij}\left(d_i-d_j\right)^2$. Noting 
that the denominator measures the variation in the degree between 
neighboring nodes on the neutral network, we may define, analogous to 
$\text{Var}(d)$, the local variation in neutrality 
$\text{LVar}(d) \equiv (2E)^{-1} \sum_{i,j}G_{ij}\left(d_i-d_j\right)^2$, where
the normalizing factor of $E$ appears because the sum is a sum over the 
edge set of the graph, and the factor of $2$ prevents double counting of 
edges. We therefore get the lower bound
\be
R -1\geq \left(\frac{n}{E}\right)^2 \frac{\text{Var}(d)^2}{\text{LVar}(d)}.
\te
Thus, although fluctuating neutrality is an essential component of 
overdispersion within the neutral evolution framework, 
the extent of overdispersion further increases  
if the graph is more sparse and has {\it smaller} 
local variation in neutrality, i.e., smaller fluctuations in neutrality 
across neighboring nodes (the latter requirement was first suggested 
in a different form by Cutler \cite{cutler2000a}). 
Significantly large overdispersion is then 
easily realized in, say, a sparse neutral network with large
diameter in which 
large {\it global} fluctuation in neutrality (degree) occurs as a  
cumulative effect of small {\it local} 
fluctuations in neutrality. 

\begin{acknowledgments}
The author acknowledges useful comments from Jesse Bloom and Claus Wilke.
This research was supported in part by US National Science Foundation 
grants 
EMT 0523643 and FIBR 0527023.
\end{acknowledgments}


\end{document}